\documentclass[pra,aps,twocolumn,showpacs,10pt]{revtex4-1}
\usepackage{amsmath,amssymb,graphicx,amsthm,dsfont,bm,color,ulem}

\begin{document}

\title{Self-trapped pulsed beams with finite power in pure cubic Kerr media excited by time-diffracting, space-time beams}
\author{Miguel A. Porras}
\affiliation{Grupo de Sistemas Complejos, ETSIME, Universidad Polit\'ecnica de Madrid, Rios Rosas 21, 28003 Madrid, Spain}

\begin{abstract}
We study the nonlinear propagation of diffraction-free, space-time wave packets, also called time-diffracting beams because its spatiotemporal structure reproduces diffraction in time. We report on the spontaneous formation of propagation-invariant, spatiotemporally compressed pulsed beams carrying finite power from exciting time-diffracting Gaussian beams in media with cubic Kerr nonlinearity at powers below the critical power for collapse, and also with other collapse-arresting nonlinearities above the critical power. Their attraction property makes the experimental observation of the self-trapped pulsed beams in cubic Kerr media feasible. The structure in the temporal and transversal dimensions of the self-trapped wave packets is shown to be the same as the structure in the axial and transversal dimensions of the self-focusing and (arrested) collapse of monochromatic Gaussian beams.
\end{abstract}

\maketitle

\section{Introduction}

\noindent Many advances in linear optics open new lines of research in nonlinear optics at high light intensities. This is what happened with the discovery of Bessel and Airy beams \cite{DURNIN,SIVILOGLOU}, their generalizations to nonlinear media \cite{WULLE,LOTTI} and application in diverse areas such as in filamentation and laser-powered material processing \cite{JUKNA,POLYNKIN}. In recent years, there is a growing interest in the so-called {\it space-time beams} (ST beams) \cite{KONDAKCI,ALONSO,KONDAKCI2,KONDAKCI3,KONDAKCI4}, wave packets whose diffraction-free behavior relies on suitably coupling the spatial and temporal degrees of freedom rather than on shaping the beam profile. The needed couplings between the spatial and temporal frequencies for diffraction-free propagation at arbitrary propagation velocity in free space and in linear dispersive media are known for some decades, mainly in the context of the so-called {\it localized waves} or modes \cite{LOCALIZED,SAARI,PORRAS,PORRAS3}, but recent advances in pulse and beam shaping techniques have made possible their practical implementation using spatial light modulators and transparent transmissive phase plates \cite{KONDAKCI2,KONDAKCI3,KONDAKCI4}. On the theoretical side, the spatiotemporal shape of these pulsed beam has been shown to reproduce the axial-transversal structure of monochromatic light that experiences diffraction spreading, that is, diffraction appears to be swapped from the longitudinal direction to time \cite{PORRAS1,PORRAS2}, which is why they are also called {\it time-diffracting beams} (TD beams). This allows to use the vast knowledge about monochromatic light beams to write down simple analytical expressions of time-diffracting beams. The time-diffraction property has been dramatically demonstrated with the synthesis of non-accelerating space-time Airy beams that accelerate instead in time \cite{KONDAKCI3}. Another remarkable property is that time-diffracting beams can have arbitrary transversal profiles, e. g., Gaussian profiles, and hence they can carry finite power.

The latest experimental results in ST or TD beam generation using transparent phase plates \cite{KONDAKCI4}, opens the possibility to generate them  at high-energy levels. It is then of interest, both from fundamental and applicative points of view, to investigate on the propagation of TD beams in nonlinear media. The paraxial approximation used in \cite{PORRAS1,PORRAS2} is particularly suited to this purpose since nonlinear propagation is modeled in the vast majority of situations by nonlinear Schr\"odinger equations \cite{SULEM}. Our analysis reveals the existence of self-trapped propagation of pulsed beams of finite power in pure (cubic) Kerr media without the necessity of introducing any additional linear or nonlinear stabilizing mechanisms such as second or higher-order dispersion \cite{KOPRINKOV,FIBICH1}, Kerr saturation, higher-order self-defocusing nonlinearity or dissipation \cite{AKHMEDIEV,SEGEV,FIBICH2}. The spontaneous reshaping of TD beams into these self-trapped wave packets supports their observability in experiments with TD beams in Kerr media. The possibility of achieving stationary propagation of strongly localized waves in the transversal direction, a possibility discarded with monochromatic light in pure Kerr media \cite{SEGEV,FIBICH2}, relies on coupling the temporal and spatial degrees of freedom. This result is closely connected with similar properties of nonlinear X waves \cite{CONTI,TRAPANI,FACCIO}, since ST couplings are at work in the stationary propagation in both cases. However, nonlinear X waves, as polychromatic and nonlinear versions Bessel beams, carry infinite power, by which they can hardly be qualified as self-trapped pulsed beams. In media with more general nonlinearities, e. g., cubic-quintic, TD beams also reshape into self-trapped pulsed beams. We demonstrate that the structure in the temporal and transversal dimensions of these wave self-trapped wave packets is a self-focusing or (arrested) collapse event {\it occurring in time,} depending on the specific nonlinearities and the input TD beam power.

\begin{figure*}[t!]
\centering
\parbox{5.5cm}{\includegraphics*[width=5.5cm]{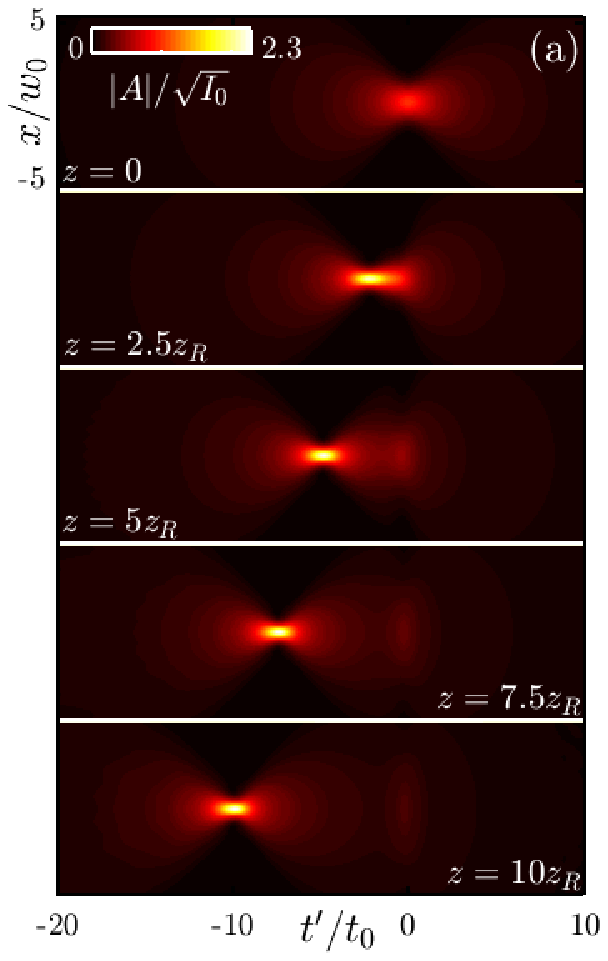}}
\parbox{5.6cm}
{\includegraphics*[width=5.6cm]{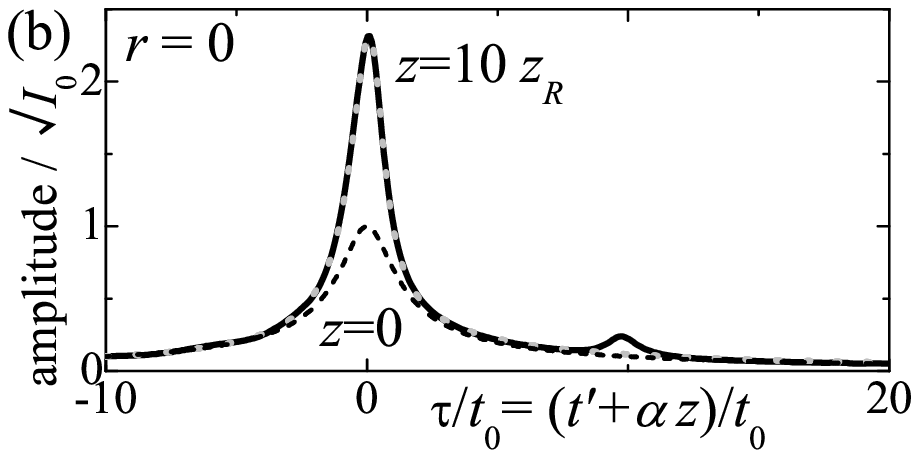}\\ \includegraphics*[width=5.6cm]{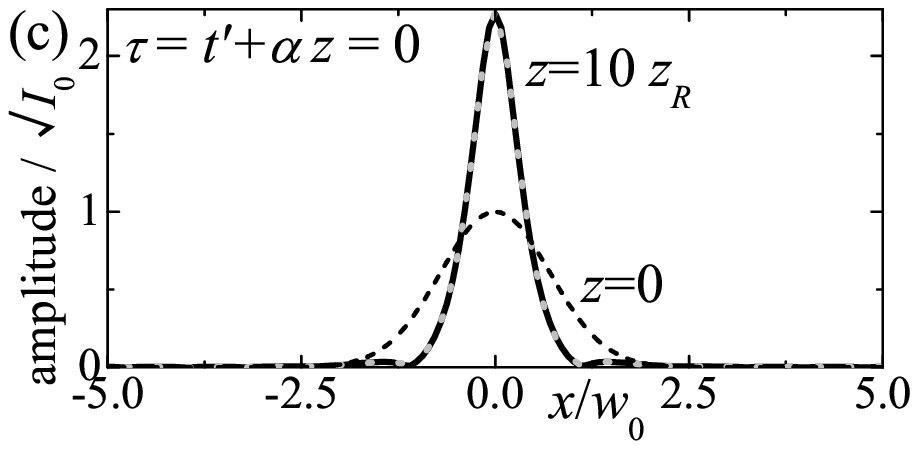}\\ \includegraphics*[width=5.6cm]{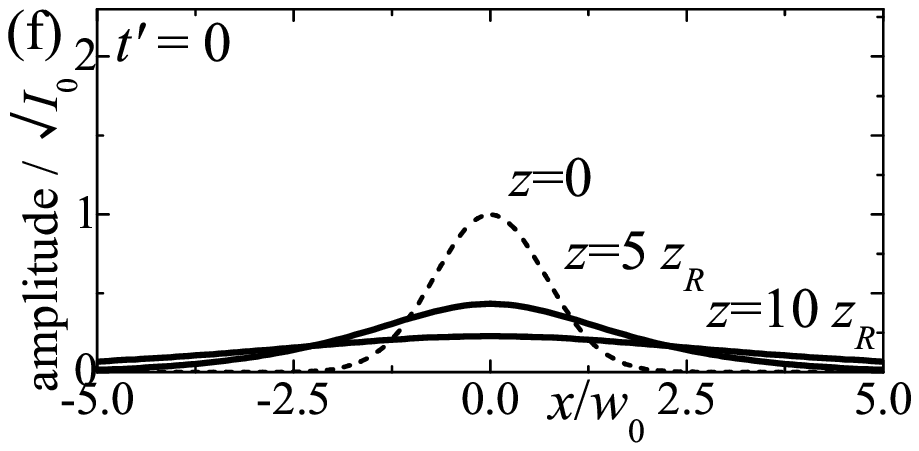}}
\parbox{5.6cm}{
\includegraphics*[width=5.6cm]{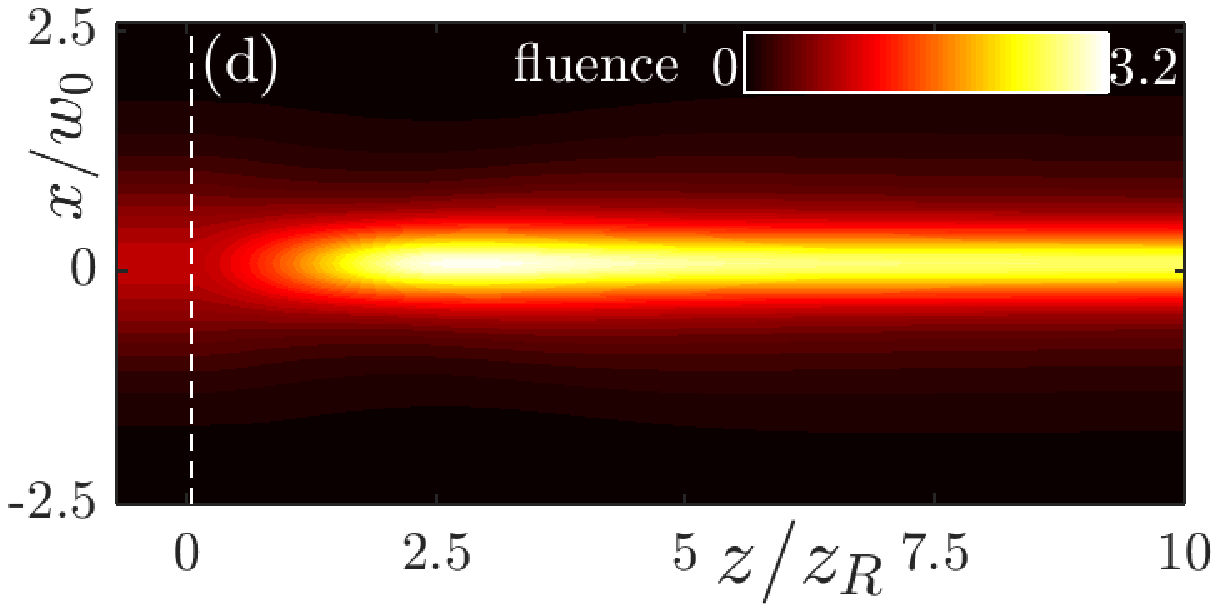}\\ \includegraphics*[width=5.6cm]{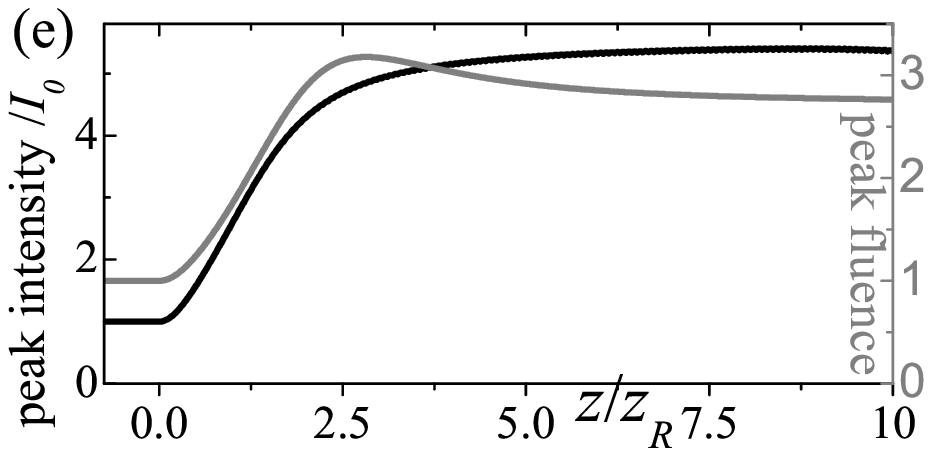}\\ \includegraphics*[width=5.6cm]{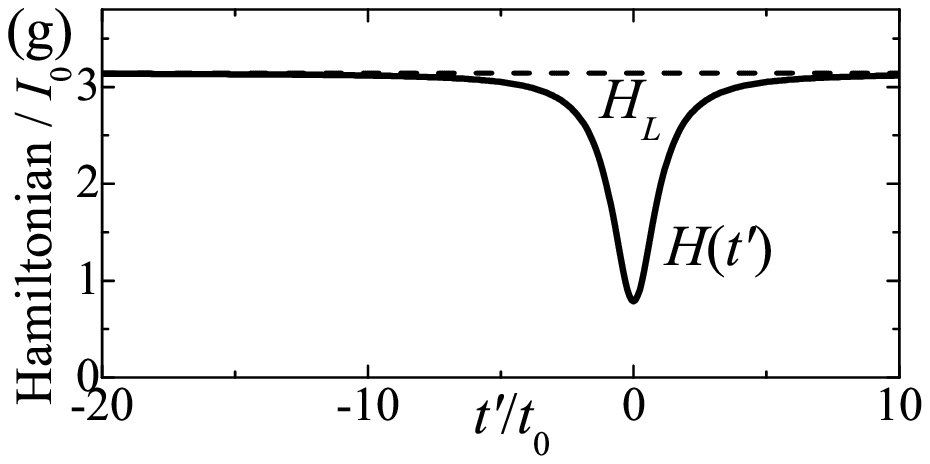}}
\caption{\label{Fig1} (a) Spatiotemporal distribution ($y=0$ section) of amplitude of a superluminal ($\alpha>0$) TDGB of power $P=0.791 P_c$ at the entrance plane $z=0$ of the nonlinear medium and as it propagates in medium. The amplitude is normalized to the amplitude $\sqrt{I_0}$ of the input TDGB. (b) On-axis amplitudes of the input TDGB (dashed curve) and of the propagation-invariant pulsed beam (solid curve) in the pulse local time $\tau=t'+\alpha z$. Gray dashed curves: On-axis amplitude for nonlinear focusing in $\tau/\alpha$ obtained as the solution of (\ref{EST}) with initial condition in (\ref{ASYMP}) at $\tau/\alpha=-20$ with the same $z_R$ and power. (c) Transversal amplitudes at $\tau=t'+\alpha z=0$ of the input TDGB (dashed curve) and of the propagation invariant pulsed beam (solid curve). Gray dashed curves: Transversal amplitude at $\tau=0$ for nonlinear focusing in $\tau/\alpha$ obtained as the solution of (\ref{EST}) with initial condition in (\ref{ASYMP}) at $\tau/\alpha=-20$ with the same $z_R$ and power. (d) Fluence of the input TDGB in free space ($z<0$) and in the nonlinear medium ($z>0$). The fluence is normalized to the peak fluence of the input TDGB. (e) Peak fluence and peak intensity at each propagation distance, normalized to their values for the input TDGB. (f) Diffracting transversal amplitude of the luminal pulse at $t'=0$ at increasing propagation distances. (g) Instantaneous linear and nonlinear Hamiltonian of the input TDGB. The instantaneous nonlinear Hamiltonian is conserved during propagation.}
\end{figure*}

\section{Time-diffracting Gaussian beams in Kerr media}

We consider the propagation in the nonlinear medium as ruled by the nonlinear Shcr\"odinger equation
\begin{equation}\label{NLSE}
\partial_z A = \frac{i}{2k_0}\Delta_\perp A + \frac{ik_0 n_2}{n_0}|A|^2 A \,,
\end{equation}
for the complex envelope $A(x,y,t',z)$ of the optical disturbance $E=Ae^{-i(\omega_0 t-k_0z)}$ of carrier frequency $\omega_0$ and propagation constant $k_0=n_0\omega_0/c$. In the above relations, $n_0$ and $n_2>0$ are the linear and nonlinear refractive indexes, $\Delta_\perp =\partial^2_x+\partial^2_y$ is the transversal Laplace operator, and $t'=t-k'_0z$, with $k'_0=dk/d\omega|_{\omega_0}$ is the local time. On using (\ref{NLSE}) we assume a paraxial regime of propagation of long enough, quasi-monochromatic pulses such that second and higher-order material dispersion plays no significant role at the propagation distances $z$ involved. Hence, no derivatives with respect to $t'$ appear in the NLSE (\ref{NLSE}) in this regime, but the envelope depends on the four variables $x,y,t',z$ for a pulsed beam.

\begin{figure*}[t!]
\centering
\includegraphics*[width=4cm]{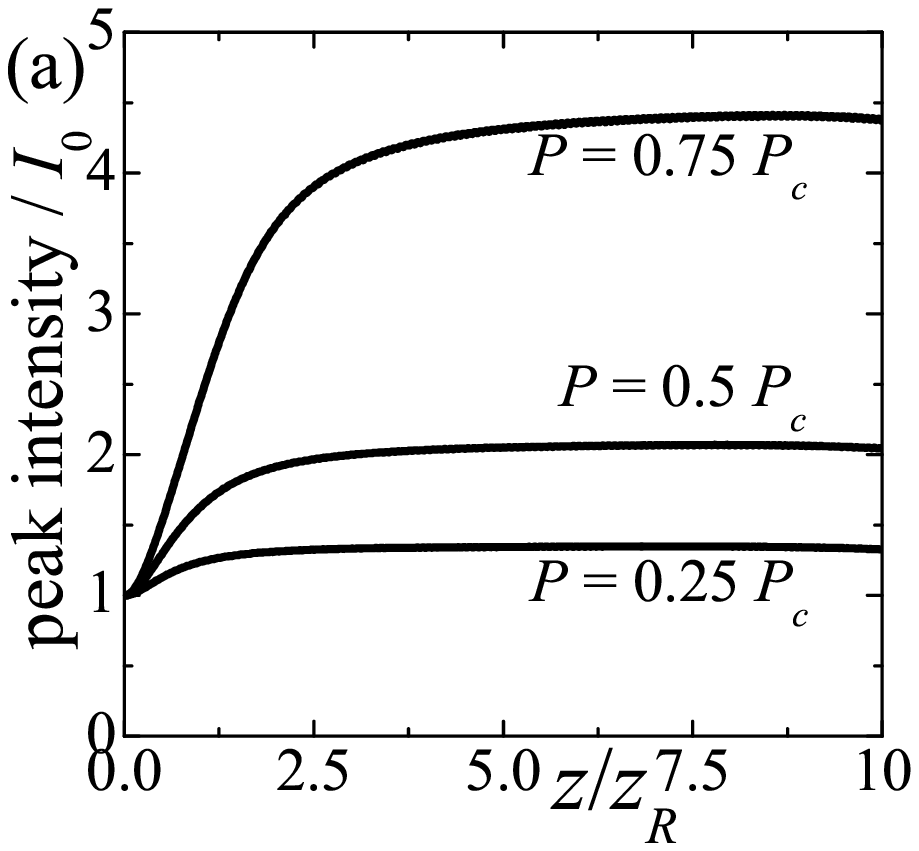}\includegraphics*[width=4cm]{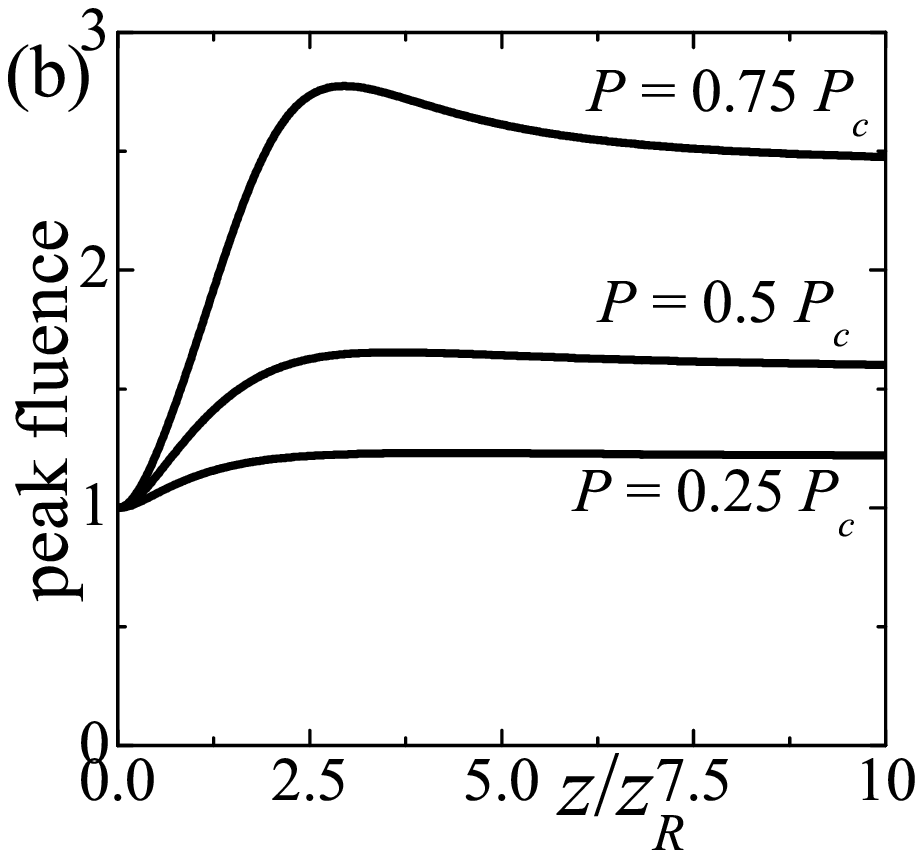}
\includegraphics*[width=4cm]{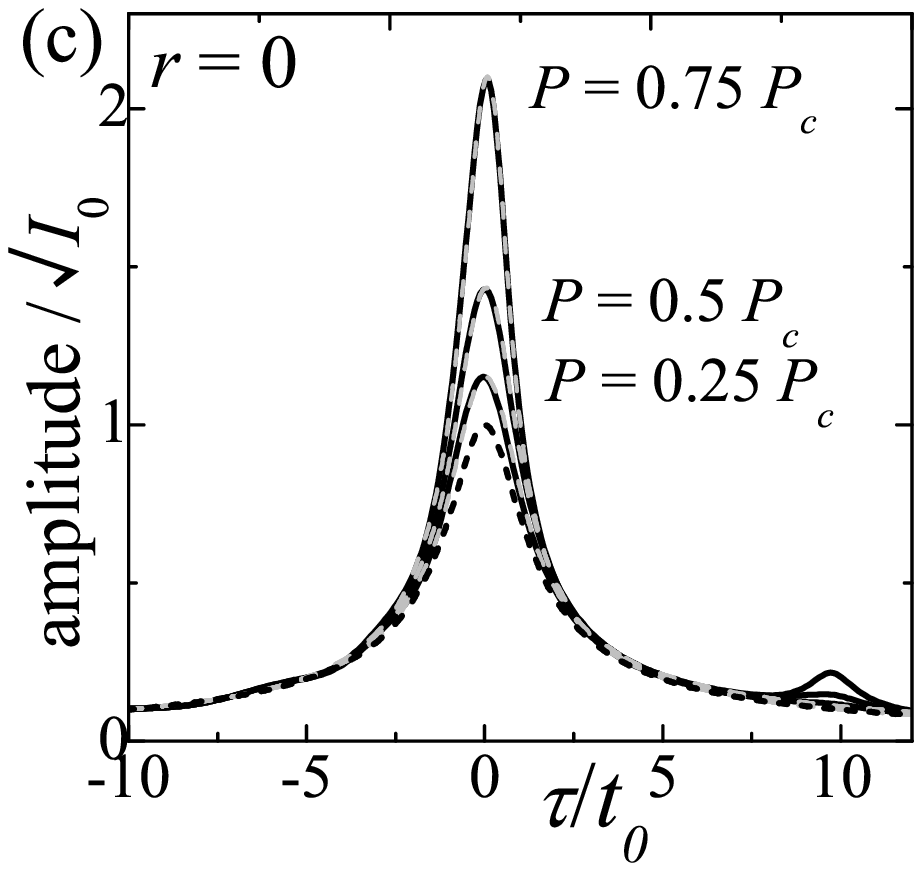}\includegraphics*[width=4cm]{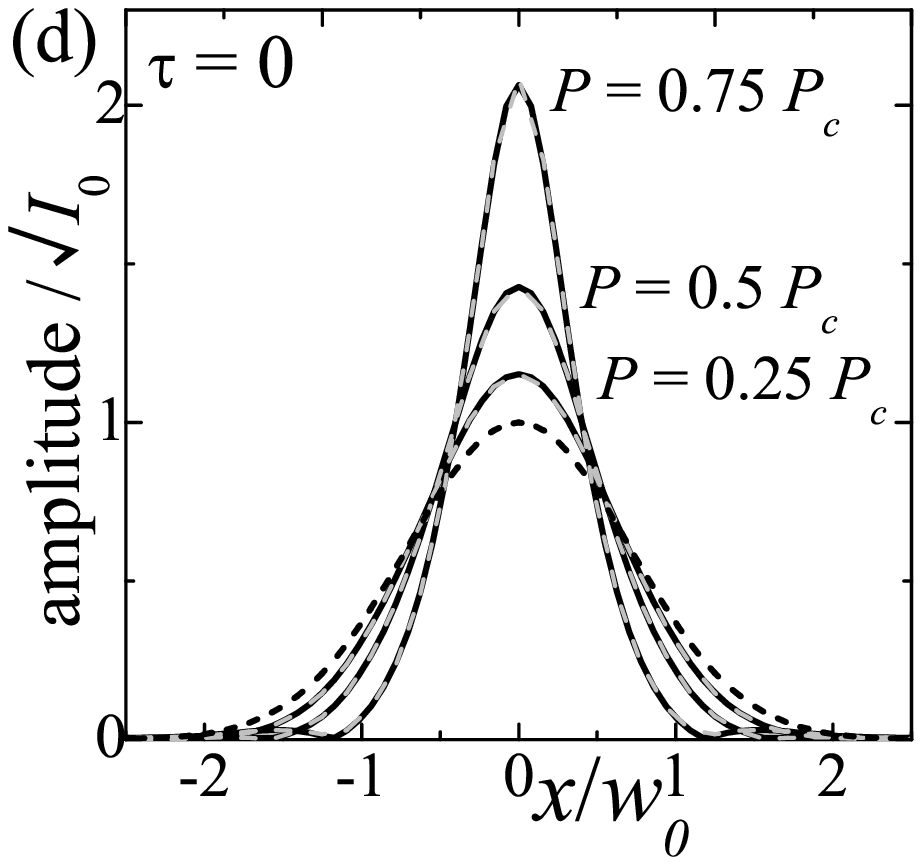}
\caption{\label{Fig2} (a) Peak intensity as function of propagation distance of input TDGBs of the indicated powers, normalized to each input peak power $I_0$. On-axis amplitudes of the input TDGB (dashed curve) and of the propagation-invariant pulsed beam (solid curve) at $z=10 z_R$ in the pulse local time $\tau=t'+\alpha z$. Gray dashed curves: On-axis amplitude for nonlinear focusing in $\tau/\alpha$ obtained as the solution of (\ref{EST}) with initial condition in (\ref{ASYMP}) at $\tau/\alpha=-20$ with the same $z_R$ and powers. (c) Transversal amplitudes at $\tau=t'+\alpha z=0$ of the input TDGB (dashed curve) and of the propagation invariant pulsed beam at $z=10 z_R$ (solid curve). Gray dashed curves: Transversal amplitude at $\tau=0$ for nonlinear focusing in $\tau/\alpha$ obtained as the solution of (\ref{EST}) with initial condition in (\ref{ASYMP}) at $\tau/\alpha=-20$ with the same $z_R$ and powers.}
\end{figure*}

Equation (\ref{NLSE}) without the nonlinear term (at low enough intensity) is satisfied by the TD Gaussian beam (TDGB) \cite{PORRAS1}
\begin{equation}\label{TDGB}
A(r,t',z)= \sqrt{I_0} \frac{-iz_R}{(t'\!+\!\alpha z)/\alpha\!-\!iz_R} \exp\left[\frac{ik_0r^2}{2\left[(t'\!+\!\alpha z)/\alpha \!-\! iz_R\right]}\right],
\end{equation}
with $\alpha\neq 0$ and $r=(x^2+y^2)^{1/2}$. Since a TDGB beam depends on $z$ only through the new local time
\begin{equation}
\tau=t'+\alpha z = t- z/v_g\,,
\end{equation}
it is a diffraction-free pulsed beam travelling undistorted at the superluminal ($\alpha>0$) or subluminal ($\alpha<0$) group velocity $1/v_g=k'_0-\alpha$, of half duration $t_0=z_R|\alpha|$, waist width $w_0^2= 2z_R/k_0$ at $\tau=0$, and peak intensity $I_0$. The TDGB can be obtained from the standard (diffracting) monochromatic Gaussian beam (MGB) $A=\sqrt{I_0}[-iz_R/(z-iz_R)]\exp[ik_0 r^2/2(z-iz_R)]$ of the same waist width $w_0$ at $z=0$ and peak intensity $I_0$ by simply replacing $z$ with $\tau/\alpha$ \cite{PORRAS2}. Consequently, the temporal-transversal structure of the TDGB is the same as the axial-transversal structure of the MGB, i. e., diffraction in time, as illustrated in Fig. \ref{Fig1}(a, top panel). If $\alpha<0$ diffraction is in addition reversed in time. The TDGB has Gaussian instantaneous transversal intensity profiles at each time. The instantaneous power $P(t')=\int |A|^2dxdy= (\pi w_0^2/2) I_0\equiv P$ is therefore finite and is independent of time, in the same way as the power of a MGB is independent of $z$, and this time-independent power $P$ is preserved on propagation along $z$. The time-independent  power reflects the weak localization of the TDGB in time, which is the same as the axial localization of a MGB that diffracts in $z$, and results in the unbounded total energy ${\cal E} =\int P(t')dt'=\infty$. For further considerations, we consider also the linear Hamiltonian \cite{FIBICH2} $H_L(t')=\int |\nabla A|^2 dx dy = \pi I_0 \equiv H_L$ that is also time-independent for TDGBs, as the $z$-independent linear Hamiltonian of a monochromatic Gaussian, and the time-independent linear Hamiltonian $H_L$ is preserved in linear propagation.

As is well-known, a MGB propagating in the nonlinear Kerr medium will collapse or not if its power is greater or smaller than the critical power $P_{c}=5.9571 n_0/k_0^2 n_2$ \cite{FIBICH3}. With monochromatic light, a propagation-invariant, self-trapped light beam has never seen to be formed because all monochromatic spatial ``solitons" in cubic Kerr media [stationary solutions of (\ref{NLSE})], including the fundamental soliton or Townes beam, high order spatial solitons, and vortex solitons, are all unstable \cite{AKHMEDIEV,SEGEV,FIBICH2}.

Within the precision of our numerical 3D+1 simulations of the NLSE (\ref{NLSE}), a TDGB introduced in the Kerr medium, e. g.,  $A(r,t',0)=\sqrt{I_0}[-iz_R/(t'/\alpha-iz_R)]\exp[ik_0 r^2/2(t'/\alpha-iz_R)]$ at the entrance plane $z=0$ of the medium, collapses, i. e., is seen to develop a singularity at a finite propagation distance, if its power is above $P_c$. With powers below $P_c$, and contrary to what happens with monochromatic light, the TDGB undergoes a transformation towards a propagation-invariant, spatiotemporally compressed and more intense pulsed beam that maintains the super or subluminal velocity of the input TDGB, as the pulsed beam centered at $t'+\alpha z=0$ in Fig. \ref{Fig1}(a) for an input superluminal TDGB ($\alpha>0$) with $P=0.791P_c$. Figs. \ref{Fig1}(b) and (c) represent, respectively, the compressed on-axis ($r=0$) temporal amplitude and the transversal amplitude at the time of arrival $t'+\alpha z=0$ at a propagation distance $z$ at which the stationary pulsed beam is already formed (solid curves), compared to the same profiles for the input TDGB (dashed curves). In Fig. \ref{Fig1}(d), the fluence of the TDGB, or time integrated intensity $F=\int |A|^2 dt'$, forming at $z<0$ what is called a ``needle of light" \cite{PORRAS2,KAMINER}, is seen to self-focus in the medium at $z>0$ and to form a narrower and more intense needle of light. The peak fluence and intensity reach $z$-independent values higher than those of the input TDGB, as in  Fig. \ref{Fig1}(e).

A finite amount of the infinite energy of the input TDGB is not coupled to the new stationary state, but forms a luminal ($v_g=1/k'_0$) diffracting pulse, [Fig. \ref{Fig1}(a) at times about $t'=0$ and Fig. \ref{Fig1}(b) at $t'+\alpha z=\alpha z$] that separates from the propagation-invariant super or subluminal pulse as they propagate. The spatiotemporal shape of the luminal pulse is that of a pulsed Gaussian beam whose transversal amplitude profile broadens on propagation [Fig. \ref{Fig1}(f)]. Then, a truly stationary state is only formed at sufficiently negative times $t'$ (for superluminal velocity) or at sufficiently positive $t'$ (for subluminal velocity).

To understand the nature of these waves we consider again the instantaneous power $P(t')=\int |A|^2 dxdy$, and the instantaneous (nonlinear) Hamiltonian \cite{FIBICH2}
\begin{equation}\label{HAMIL}
H(t')= \int |\nabla A|^2 dxdy- \frac{k_0^2 n_2}{n_0}\int |A|^4 dxdy\,,
\end{equation}
which are conserved during propagation {\it at any particular instant of time} $t'$, since $t'$ only appears parametrically in the NLSE (\ref{NLSE}), and can therefore be calculated with the input TDGB at $z=0$. Since the input TDGB has time-independent power $P(t') =(\pi w_0^2/2) I_0 \equiv P$, the same is true for the propagated pulse, including the negative or positive times with the super or subluminal diffraction-free pulse and the times about zero with the luminal pulsed Gaussian beam. The conserved instantaneous Hamiltonian, calculated from (\ref{HAMIL}) with the input TDGB, is given by
\begin{equation}\label{HAMIL2}
H(t')=H_L \left[1-0.948 \frac{P}{P_c}\frac{1}{1+(t'/t_0)^2}\right] \, ,
\end{equation}
and is represented in Fig. \ref{Fig1}(f) for $P=0.791 P_c$. Thus, the luminal pulsed Gaussian beam diffracts because the Hamiltonian about $t'=0$ is sufficiently positive (its power is below $P_c$), and the super or subluminal propagation-invariant pulse must be characterized by a time-independent power $P$ equal to that of the input TDGB, and a time-independent nonlinear Hamiltonian equal the linear Hamiltonian $H_L$ of the input TDGB.

\section{Self-trapped pulsed beams that self-focus and collapse in time}

The above analysis suggests to search for stationary solutions of the NLSE (\ref{NLSE}) of the form $A=A(x,y,t'+\alpha z)= A_\alpha(x,y,\tau)$, travelling undistorted at the group velocity $1/v_g=k'_0-\alpha$. With this ansatz, the NLSE (\ref{NLSE}) yields
\begin{equation}\label{EST}
\partial_\tau A_\alpha = \frac{i}{2k_0\alpha}\Delta_\perp A_\alpha + \frac{i k_0n_2}{n_0\alpha}|A_\alpha|^2 A_\alpha\,,
\end{equation}
which is formally the same as (\ref{NLSE}) with $z$ replaced with the length $\tau/\alpha$ proportional to time, and for a complex envelope that depends on the three variables $x,y,\tau$ instead of the four variables $x,y,t',z$. Equation (\ref{EST}) then describes the same dynamics as that of a strictly monochromatic light beam along $z$ under the action of diffraction and self-focusing, but this dynamics is displayed in $\tau/\alpha$, and represents the spatiotemporal shape of a propagation-invariant pulsed beam. The power $P=\int |A_\alpha|^2 dxdy$ and the Hamiltonian,
\begin{equation}\label{HAMILEST}
H=\int  |\nabla A_\alpha|^2 dxdy- \frac{k_0^2 n_2}{n_0}\int |A_\alpha|^4 dxdy \,,
\end{equation}
of the solutions of (\ref{EST}) are, as desired, independent of $\tau$, in the same way as for monochromatic light they are independent of $z$. Separable solutions of (\ref{EST}) in $r$ and $\tau$ are readily seen to be (using separation of variables) necessarily of the form $A_\alpha=e^{i\Omega\tau}g(r)$, with $\Omega$ a small frequency shift and $g(r)$ the fundamental Townes, high-order, or vortex ``solitons". The propagation-invariant pulsed beam in Fig. \ref{Fig1} is, however, obviously nonseparable in $r$ and $\tau$. These nonseparable solutions have not been investigated with this meaning, i. e., as describing the spatiotemporal shape propagation-invariant pulses.

\begin{figure*}[t]
\centering
\parbox{6cm}{\includegraphics*[width=6cm]{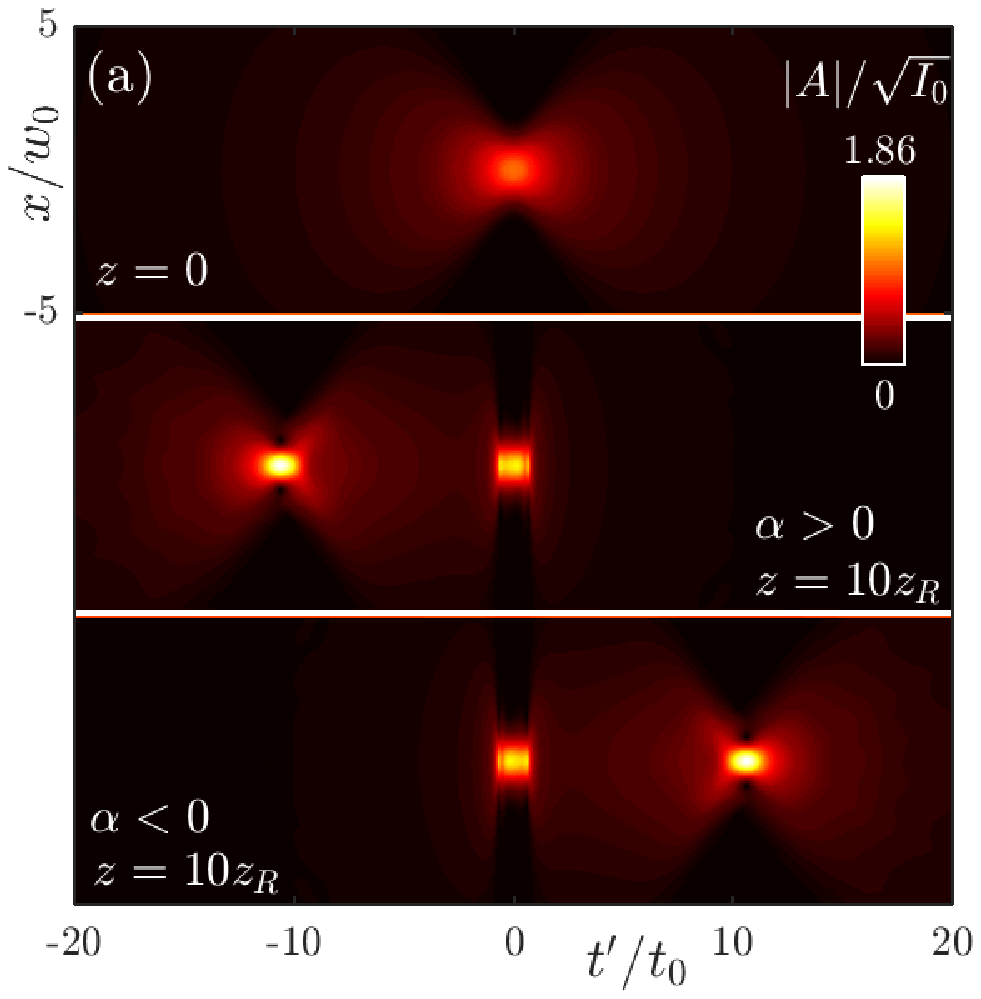}}
\parbox{5.6cm}{\includegraphics*[width=5.6cm]{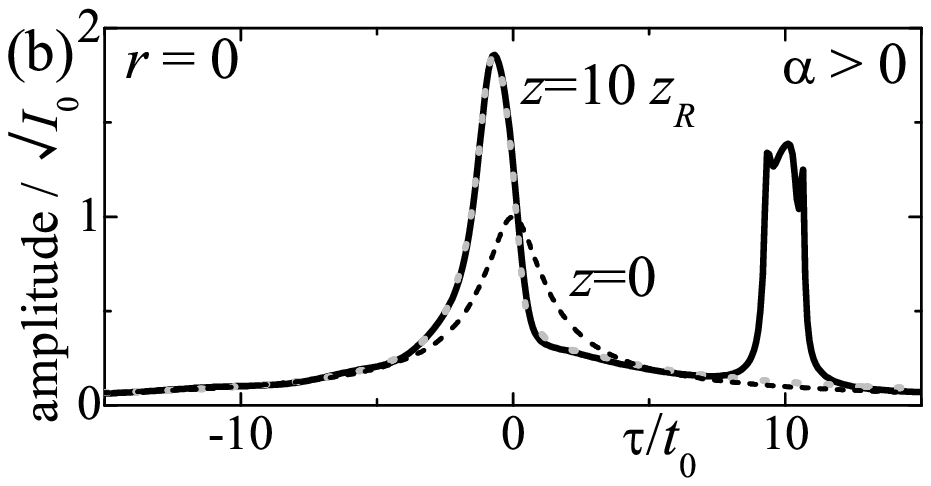} \\ \includegraphics*[width=5.6cm]{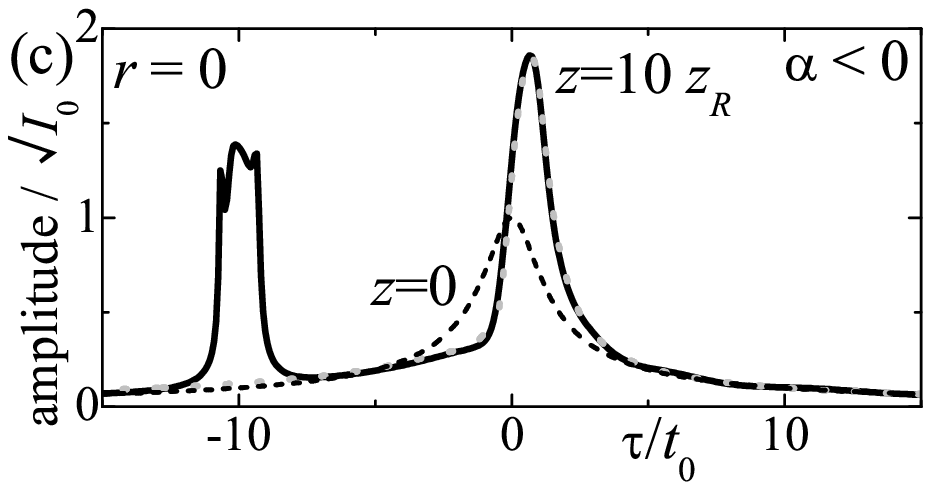}}
\parbox{5.6cm}{\includegraphics*[width=5.6cm]{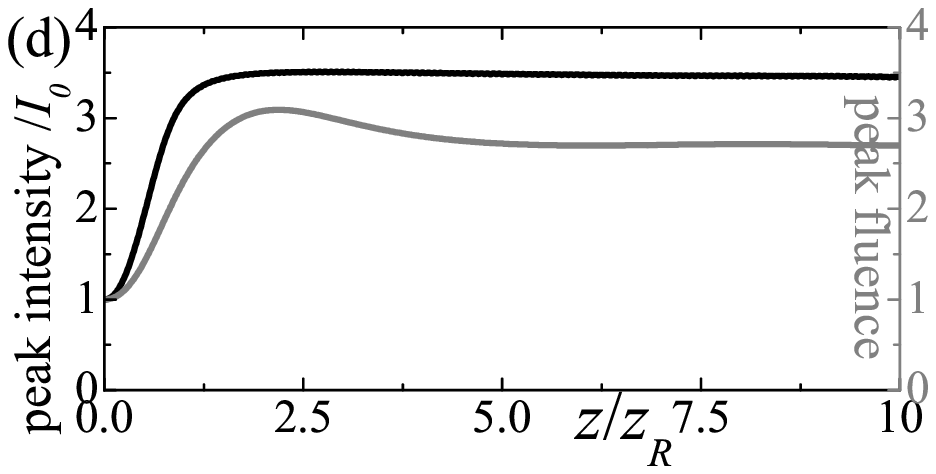} \\ \includegraphics*[width=5.6cm]{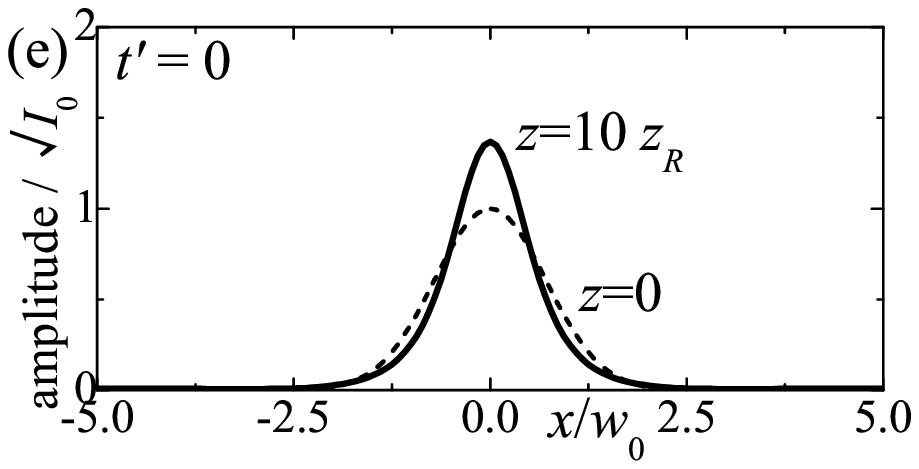}}
\caption{\label{Fig3} (a) For input super and subluminal TDGBs of power $P=1.5 P_c$ in a medium with self-focusing cubic and self-defocusing quintic nonlinearities ($n=n_0 +n_2 I + n_4 I^2$) such that $|n_4 I_0^2/n_2 I_0|=0.2$, spatiotemporal distribution of amplitude ($y=0$ section) at $z=0$ and at sufficiently long distance ($z=10 z_R$). (b) and (c) On-axis amplitude of the input TDGB (dashed curves) and of the propagation-invariant pulsed beam (solid curves) as functions of the local time $\tau$ in the respective cases of super and subluminal input TDGB. Gray dashed curves: On-axis amplitude for nonlinear focusing along the length $\tau/\alpha$ of the initial condition in (\ref{ASYMP}) at $\tau/\alpha=-20$ with the same $z_R$ and power in the same cubic-quintic medium. (d) Peak fluence and peak intensity at each propagation distance, normalized to their values for the input TDGB. (e) Transversal amplitude profile of the luminal spatial soliton formed about $t'=0$.}
\end{figure*}

We consider the solution to (\ref{EST}) that behaves asymptotically at $\tau/\alpha\rightarrow -\infty$ ($\tau\rightarrow -\infty$ for $\alpha>0$ and $\tau\rightarrow +\infty$ for $\alpha<0$), times at which the intensity is sufficiently low, as
\begin{equation}\label{ASYMP}
A_\alpha(r,\tau)\stackrel{\tau/\alpha\rightarrow -\infty}{\longrightarrow} \sqrt{I_0}\frac{-iz_R}{\tau/\alpha-iz_R} \exp\left[\frac{ik_0r^2}{2\left[\tau/\alpha - iz_R\right]}\right],
\end{equation}
i. e., as the input TDGB with the same $z_R$, $\alpha$ and $I_0$ at these long times. Since the power is constant in time, it is thus fixed to the power $P$ of the input TDGB. Also, since the nonlinear term of the Hamiltonian in (\ref{HAMILEST}) is negligible at $\tau/\alpha\rightarrow -\infty$, the Hamiltonian is $H\stackrel{\tau/\alpha\rightarrow -\infty}{\longrightarrow} H_L$ asymptotically, and being constant in time, the Hamiltonian is $H=H_L$ at any time. In the length $\tau/\alpha$, this solution represent the focusing of a sufficiently broad and converging MGB to a linear focus at $\tau/\alpha=0$ of half-depth $z_R$, assisted by self-focusing. If $P>P_c$, as is well-known, the solution collapses, but if $P<P_c$, it does not. In the example with $\alpha>0$ and $P=0.791 P_c$, the gray dashed curves in Figs. \ref{Fig1} (b) and (c) represent the on-axis amplitude and the transversal amplitude at $\tau=0$ obtained by solving numerically (\ref{EST}) taking as the initial condition the right hand side of (\ref{ASYMP}) at $\tau/\alpha=-\tau_{in}/\alpha = -20 \ll 0$, so that the pulse focuses linearly at these initial times. The complete solution, except at the times where the luminal diffracting pulse is located, matches accurately with the spatiotemporal shape of the propagation-invariant superluminal pulsed beam excited by the input TDGB. The spatiotemporal structure of the propagation-invariant pulsed beam can then be said to be that of  self-focusing in time.

We confirm this result in Fig. \ref{Fig2} for input TDGBs with different subcritical powers. In all cases the peak intensity and the fluence stabilize with propagation distance at constant values [Figs. \ref{Fig2} (a) and (b)]. The propagation-invariant on-axis amplitude and the transversal profile at the pulse center, $\tau=0$ at these distances are depicted in Figs. \ref{Fig2}(c) and (d) as solid black curves. They are indistinguishable at the scale of the figures, except for the uncoupled diffracting pulsed Gaussian beam at $\tau=\alpha z$, from the dashed gray curves, obtained by solving (\ref{EST}) with the initial condition (\ref{ASYMP}) at sufficiently negative $\tau/\alpha$ with same $z_R$, $\alpha$ and respective powers.

We have restricted ourselves to the cubic Kerr nonlinearity, and to subcritical powers to avoid the formation of a singularity, because self-trapped propagation at finite power has not previously been reported with this nonlinearity, to the author knowledge. However the same result holds with other nonlinearities, and in particular with nonlinearities that arrest the collapse induced by the cubic nonlinearity at supercritical powers. For example, Fig. \ref{Fig3}(a) displays the spatiotemporal amplitude distribution at sufficiently long distance (lower panels) obtained by solving the NLSE (\ref{NLSE}) with the additional quintic term $(ik_0 n_4/n_0)|A|^4 A$, with $n_4<0$. The initial conditions at $z=$ are superluminal ($\alpha>0$) and subluminal ($\alpha<0$) TDGBs (top panel) of power $P=1.5 P_c$, and the strength of the quintic term is determined by $|n_4 I_0^2/n_2 I_0|=0.2$. The on-axis amplitudes at long distances as functions of time are represented in Fig. \ref{Fig3}(b) and (c) as solid black curves. Clearly, the only difference between the final stationary states with input superluminal and subluminal TDGBs is a time reversal. The peak intensity and fluence are seen in Fig. \ref{Fig3}(d) to remain constant at these distances. The energy that is not coupled to the superluminal or subluminal pulsed beam forms now a (slightly vibrating) luminal spatial soliton (as supported by the cubic-quintic model \cite{FIBICH2}), whose transversal amplitude profile is displayed in Fig. \ref{Fig3}(e). The dashed gray curves in Figs. \ref{Fig3}(b) and (c) represent the on-axis amplitude obtained by solving numerically (\ref{EST}) with the additional term $(ik_0 n_4/n_0\alpha)|A|^4 A$, using as initial conditions the right hand side of (\ref{ASYMP}) at $\tau/\alpha=-\tau_{in}/\alpha = -20 \ll 0$ (large negative or positive times for super or subluminal input TDGB). The spatiotemporal structure of the self-trapped superluminal pulsed beam is then described by collapse arrested by the self-defocusing quintic nonlinearity displayed in time, and temporally reversed in the subluminal case.

\section{Conclusion}

In conclusion, we have shown that it is possible to observe self-trapped pulsed beam propagation in cubic Kerr media at finite power below the critical power for collapse, and also above the critical power in non-Kerr media. These pulsed beams are excited by TD beams, which can be generated  using recent techniques for beam and pulse shaping to realize the required spatiotemporal correlations \cite{KONDAKCI2,KONDAKCI3,KONDAKCI4}, or, as proposed in \cite{PORRAS2}, in group mismatched second-harmonic generation. We have limited the analysis to input TDGB for conciseness, but similar results are expected to hold with other TD beams, since any monochromatic light beam has a TD counterpart and the self-focusing mechanism is the same. For example, at sufficiently higher powers ($P\simeq 10 P_c$ or higher) and with super-Gaussian TD beams, one would expect to observe a propagation-invariant pulsed beam with the temporal shape of a self-focusing ring that breaks into multiple temporal filaments, as described in standard self-focusing theory \cite{FIBICH4}. The existence of these waves opens up new possibilities in nonlinear optics applications such as filamentation and material processing, particularly in relation to their strong transversal localization and not requiring materials with tailored nonlinearities to support the self-trapped propagation.

The author acknowledges support from Projects of the Spanish Ministerio de Econom\'{\i}a y Competitividad No. MTM2015-63914-P and No. FIS2017-87360-P.

\end{document}